\documentstyle[12pt]{article}
\begin{document}

\begin{center}
{\Large \bf ``CLASSICAL'' PROPAGATOR AND PATH INTEGRAL IN THE PROBABILITY
REPRESENTATION OF QUANTUM MECHANICS}

\end{center}

\bigskip

\begin{center}
{\bf Olga Man'ko~\footnote{Assosiate Member of ICTP (Trieste).}
 and V. I. Man'ko }  
\end{center}
\medskip
\begin{center}
{\it 
P.\,N.\,Lebedev Physical Institute, Russian Academy of Sciences,
Leninskii Pr. 53, Moscow 117924, Russia} \\
e--mails:~~~omanko@sci.lebedev.ru~~~~manko@na.infn.it
\end{center}

\begin{abstract}
\noindent 
In the probability representation of the standard quantum mechanics,
the explicit expression (and its quasiclassical van-Fleck approximation)
for the ``classical'' propagator (transition probability distribution),
which completely describes the quantum system's evolution, is found in 
terms of the quantum propagator. An expression for the ``classical''
propagator in terms of path integral is derived. Examples of free motion
and harmonic oscillator are considered. The evolution equation in the
Bargmann representation of the optical tomography approach is obtained.
\end{abstract}

\bigskip

\section{Introduction}

\noindent

In~\cite{ManciniPL,ManciniFP}, the new formulation of quantum mechanics,
in which a quantum state was described by the tomographic probability
distribution (called ``marginal distribution'') instead of density matrix
or the Wigner function was suggested and the quantum evolution
equation of the generalized ``classical'' Fokker--Planck-type equation 
alternative to the Schr\"odinger equation was found. 
The physical meaning of the marginal distribution was elucidated as the
probability distribution for the position measured in an ensemble of
scaled and rotated reference frames in the classical phase space of the
system under consideration. The ``classical'' propagator describing the 
transition probability from an initial position labeled by parameters
of the initial reference frame of the ensemble to a final position lebeled
by parameters of the final reference frame of the ensemble was 
introduced~\cite{ManciniFP}.

The quantum propagator (Green function of the evolution equation for the 
density matrix) was expressed in terms of the ``classical'' propagator 
in~\cite{OlgaVolJRLR97}. Properties of the ``classical'' propagator and its
relation to quantum time-dependent integrals of motion~[4--9]
were studied
in~\cite{RosaPhR98}. An analog of the Schr\"odinger equation for energy 
levels in terms of the marginal distribution was discussed in~\cite{ManJRLR96}.
Different examples like quantum dissipation and quantum top were considered
in~\cite{SafTMF,SafYadF}. An extension of the new formulation of quantum
mechanics to the case of spin was given in~[14--17].
An example of quantum diffraction in time~\cite{MoshPhR52} in the framework 
of the new (probability) representation of quantum mechanics was considered
recently~\cite{Sharma}.

Feynman suggested the formulation of quantum mechanics by means of the path
integral method~\cite{Feynman}. In his formulation, the quantum 
transition-probability amplitude (Green function of the Schr\"odinger
evolution equation for the wave function) is expressed as path integral
determined by the classical action of the system under consideration.
To use the path integral method in the new (probability) representation 
of quantum mechanics, one needs to have a formula for the ``classical''
propagator in terms of the quantum propagator, which is inverse of the
formula for the quantum propagator in terms of the ``classical'' propagator
derived in~\cite{OlgaVolJRLR97}; till now, this formula was not available.

The aim of this paper is to derive the explicit relationship of the 
``classical'' propagator (Green function of the evolution equation 
for marginal distribution) in terms of Green function of the 
Schr\"odinger evolution equation. 
Another goal of this study is to express, 
in view of the relationship obtained, 
the ``classical'' propagator  as path integral determined 
by the classical action.

\section{Marginal Distribution and Wave Function}

\noindent

 Let a quantum state be described by the wave function 
$\Psi\left(x\right)$.
The nonnegative marginal distribution
$w\left(X,\,\mu,\,\nu\right),$
which desribes the quantum state, is given by the
relationship~\cite{Mendes,Sharma}
\begin{equation}\label{Path1}
w\left(X,\mu,\nu\right)=\frac {1}{2\,\pi\,|\nu|}\left|\int
\Psi\left(y\right)\exp\left(\frac {i\,\mu}{2\,\nu}\,y^2
-\frac {iX}{\nu}\,y\right)\,dy\right|^2,
\end{equation}
where the random coordinate $X$ corresponds to the particle's position
and the real parameters $\mu$ and $\nu$ label the reference frame in 
the classical phase space, in which the position is measured. The density
matrix  of the pure state 
$\rho_\Psi\left(x,\,x'\right)$
in the position representation is expressed in terms of the marginal
distribution (see, for example,~\cite{RosaPhL98}\,)
\begin{eqnarray}\label{Path2}
\rho_\Psi\left(x,x'\right)&=&\Psi\left(x\right)\,\Psi^*
\left(x'\right)\nonumber\\
&=&\frac {1}{2\,\pi}\int w\left(X,\mu,x-x'\right)\,
\exp\left[i\left(X-\mu\,\frac {x+x'}{2}\right)\right]\,d\mu\,dX\,.
\end{eqnarray}
Thus, if one knows the density matrix of the pure state (or wave function),
the marginal distribution is also known, in view of Eq.~(\ref{Path1}).
Correspondingly, if one knows the marginal distribution, the density matrix
is also known, in view of the inverse relationship~(\ref{Path2}).

\section{``Classical'' Propagator}

\noindent

 The evolution of the marginal distribution 
$w\left(X,\,\mu,\,\nu,\,t\right)$
can be described by means of the ``classical'' propagator
$\Pi\left(X_2,\,\mu_2,\,\nu_2,\,X_1,\,\mu_1,\,\nu_1,\,t_2,\,t_1\right)$,
in view of the integral relationship~\cite{ManciniFP}
\begin{equation}\label{Path3}
w\left(X_2,\mu_2,\nu_2,t_2\right)=\int
\Pi\left(X_2,\mu_2,\nu_2,X_1,\mu_1,\nu_1,t_2,t_1\right)
w\left(X_1,\mu_1,\nu_1,t_1\right)\,dX_1\,d\mu_1\,d\nu_1\,.
\end{equation}
Below, we will use also the notation
$\Pi\left(X_2,\,\mu_2,\,\nu_2,\,X_1,\,\mu_1,\,\nu_1,\,t\right)$
for the ``classical'' propagator in the case of $\,t_1=0,~~t_2=t.$ 

Having in mind the physical meaning of the ``classical'' propagator, we know
that it must satisfy  the nonlinear relationship
\begin{eqnarray}\label{Path4}
\Pi\left(X_3,\mu_3,\nu_3,X_1,\mu_1,\nu_1,t_3,t_1\right)
&=&\int\Pi\left(X_3,\mu_3,\nu_3,X_2,\mu_2,\nu_2,t_3,t_2\right)\nonumber\\
&&\otimes\Pi\left(X_2,\mu_2,\nu_2,X_1,\mu_1,\nu_1,t_2,t_1\right)
\,dX_2\,d\mu_2\,d\nu_2\,.
\end{eqnarray}
Formula~(\ref{Path4}) describes the obvious standard property of the 
transition probability from an initial point $X_1$ to a final point $X_3$
via an intermediate point $X_2.$ The only pecularity of this formula consists
in the fact that the initial, final, and intermediate points 
$\,X_1,\,X_3,\,X_2\,$ are considered in their own reference frames labeled
by the paramaters 
$\,\mu_1;\,\nu_1,\,$
$\,\mu_3;\,\nu_3,\,$ and
$\,\mu_2;\,\nu_2,\,$ respectively.
Empoying this property one produces integration in Eq.~(\ref{Path4}) not 
only over points $X_2$ but also over the reference frames' parameters.

In view of Eq.~(\ref{Path1}), the marginal distribution
$w\left(X,\,\mu,\,\nu\right)$
has the property~\cite{RosaPhL98}
\begin{equation}\label{Path5}
w\left(aX,a\mu,a\nu\right)=\frac {1}{|a|}
\,w\left(X,\mu,\nu\right).
\end{equation}
Due to this, the ``classical'' propagator has the analogous property,
\begin{equation}\label{Path6}
\Pi\left(bX,b\mu,b\nu,bX',b\mu',b\nu',t\right)=\frac {1}{|b|^3}
\Pi\left(X,\mu,\nu,X',\mu',\nu',t\right).
\end{equation}
Equation~(\ref{Path6}) provides the connection of two Fourier components, 
namely,
\begin{equation}\label{Path7}
\Pi_{\rm F}\left(1,\mu,\nu,X',\mu',\nu',t\right)=\int
\Pi\left(X,\mu,\nu,X',\mu',\nu',t\right)e^{iX}\,dX
\end{equation}
and
\begin{equation}\label{Path8}
\Pi_{\rm F}\left(k,\mu,\nu,X',\mu',\nu',t\right)=\int
\Pi\left(X,\mu,\nu,X',\mu',\nu',t\right)e^{ikX}\,dX\,.
\end{equation}
As a result, one has
\begin{equation}\label{Path9}
\Pi_{\rm F}\left(k,\mu,\nu,X',\mu',\nu',t\right)=k^2\,
\Pi_{\rm F}\left(1,k\mu,k\nu,kX',k\mu',k\nu',t\right).
\end{equation}

\section{Green Function and ``Classical'' Propagator}

\noindent

 The Green function 
$G\left(x,\,y,\,t\right)$
(quantum propagator) is determined by the relationship
\begin{equation}\label{Path10}
\Psi\left(x,t\right)=\int 
G\left(x,y,t\right)
\Psi\left(y,\,t=0\right)\,dy\,.
\end{equation}
Elaborating Eqs.~(\ref{Path2}) and (\ref{Path3}) for 
arbitrary time $t,$ one arrives at
\begin{eqnarray}\label{Path11}
\int G\left(x,y,t\right)G^*\left(x',z,t\right)
\Psi\left(y,\,t=0\right)\,\Psi^*\left(z,\,t=0\right)\,dy\,dz
&=&\frac {1}{2\,\pi}\int \,dX'\,d\mu'\,d\nu'dX\,d\mu\,
~w\left(X',\mu',\nu'\right)\nonumber\\
&&\otimes\Pi\left(X,\mu,x-x',X',\mu',\nu',t\right)\nonumber\\
&&\otimes
\exp\left[i\left(X-\mu\,\frac {x+x'}{2}\right)\right].
\end{eqnarray}
For the wave function in~(\ref{Path11}), we used the notation
$$\Psi\left(y\right)\equiv \Psi\left(y,\,t=0\right).$$
Taking into account relationship~(\ref{Path2}) for the product
$\Psi\left(y,\,t=0\right)\,\Psi^*\left(z,\,t=0\right)$
and the relationship
$$w\left(X',\mu',y-z\right)=\int w\left(X',\mu',\nu'\right)\,
\delta\left(y-z-\nu'\right)\,d\nu',$$
in view of relationship~(\ref{Path9}) between the Fourier components,
one obtains
\begin{eqnarray}\label{Path12}
\Pi\left(X,\mu,\nu,X',\mu',\nu',t\right)&=&\frac {1}{4\,\pi^2}\int k^2\,
G\left(a+\frac {k\nu}{2},y,t\right)\,G^*\left(a-\frac {k\nu}{2},z,t\right)\,
\delta\left(y-z-k\nu'\right)\nonumber\\
&&\otimes\exp\left[ik\left(X'-X+\mu a-\mu'\,\frac {y+z}{2}\right)\right]
\,dk\,dy\,dz\,da\,.
\end{eqnarray}
Equation~(\ref{Path12}) can be rewritten in the form
\begin{eqnarray}\label{Path13}
\Pi\left(X,\mu,\nu,X',\mu',\nu',t\right)&=&\frac {1}{4\,\pi^2}\int k^2\,
G\left(a+\frac {k\nu}{2},z+k\nu',t\right)\,
G^*\left(a-\frac {k\nu}{2},z,t\right)\nonumber\\
&&\otimes\exp\left[ik\left(X'-X-\frac {k\mu'\nu'}{2}-\mu' z+\mu a\right)
\right]\,dk\,dz\,da\,.
\end{eqnarray}
Relationships~(\ref{Path12}) and (\ref{Path13}) provide the expression for
the ``classical'' propagator in terms of Green function of the Schr\"odinger
evolution equation. One can see that the ``classical'' propagator depends
on the difference of positions $X-X'$.

\section{``Classical'' Propagator and Path Integral}

\noindent

 The quantum propagator (Green function of the Schr\"odinger evolution
equation) 
$G\left(x_2,\,x_1,\,t_2,\,t_1\right)$ 
can be presented as path integral~\cite{Feynman}
\begin{equation}\label{Path14}
G\left(x_2,x_1,t_2,t_1\right)
=\int_{\left(x_1,\,t_1\right)}^{\left(x_2,\,t_2\right)}\exp\left\{i
S\left[x\left(t\right)\right]
\right\}\,{\cal D}\left[x\left(t\right)\right],
\qquad \hbar=1\,,
\end{equation}
where the functional $S\left[x\left(t\right)\right]$ is 
the classical action
\begin{equation}\label{Path15}
S\left[x\left(t\right)\right]=\int_{t_1}^{t_2}
L\Big(x(t),\,\dot x(t),\,t\Big)\,dt\,,
\end{equation}
with Lagrangian $L\Big(x(t),\,\dot x(t),\,t\Big)$
 of the form $\left(m=0\right)$
\begin{equation}\label{Path16}
L\Big(x(t),\,\dot x(t),\,t\Big)=\frac {\dot x^2}{2}-U\left(x,t\right).
\end{equation}
The classical trajectory is given by the minimal action principle
$$\delta S=0\,,$$
that is equivalent to the Lagrange--Euler equation
\begin{equation}\label{Path17}
\frac {d}{dt}\,
\frac {\partial L}{\partial \dot x}
=\frac {\partial L}{\partial x}\,.
\end{equation}
The action functional can be considered as a function of several variables
\begin{equation}\label{Path18}
S\left[x\left(t\right)\right]\approx \sum_{n=1}^N\left[
\frac{\left(x_n-x_{n-1}\right)^2}{2\,\Delta t}-U\left(x_n,\,t_n\right)
\,\Delta t\right],
\end{equation}
where 
$$x_n\equiv x\left(t_n\right),\qquad t_2-t_1=N\Delta t\,,
\qquad t_{n+1}-t_n=\Delta t\,.$$

Thus, path integral can be approximately considered as multidimensional
integral over the variables\\
 $x_1,\,x_2,\,\ldots\,,\,x_N.$

In view of Eq.~(\ref{Path14}), relation~(\ref{Path13}) can be
rewritten as the path integral representation for the ``classical''
propagator
\begin{eqnarray}\label{Path19}
\Pi\left(X,\mu,\nu,X',\mu',\nu',t\right)&
=&\frac {1}{4\,\pi^2}\int
k^2\int _{\left(z+k\nu',\,0\right)}
^{\left(a+k\nu/2,\,t\right)}
~~\exp\left\{iS\left[x(t)\right]\right\}\,{\cal D}\left[x(t)\right]
\nonumber\\[2mm]
&&\otimes \int_{\left(z,\,0\right)}^{\left(a-k\nu/2,\,t\right)}
~~\exp\left\{-iS\left[y(t)\right]\right\}\,{\cal D}\left[y(t)\right]
\nonumber\\[2mm]
&&\otimes\exp\left[ik\left(X'-X-\frac {k\mu'\nu'}{2}-\mu' z+\mu a\right)
\right]\,dk\,dz\,da\,.
\end{eqnarray}

\section{Free Motion and Harmonic Oscillator}

\noindent

 For a free particle $\left(\hbar=m=1\right),$ Green function is
of the form
\begin{equation}\label{Path20}
G\left(x,y,t\right)=\frac {1}{\sqrt {2\,\pi i t}}\,\exp\left[\frac 
{i\left(x-y\right)^2}{2t}\right].
\end{equation}
The ``classical'' propagator for free motion found in~\cite{OlgaVolJRLR97}
has the appearance
\begin{equation}\label{Path21}
\Pi_{\rm f}\left(X,\mu,\nu,X',\mu',\nu',t\right)=\delta\left(X-X'\right)
\,\delta\left(\mu -\mu'\right)\,\delta\left(\nu-\nu'+\mu t\right).
\end{equation}
We can calculate the ``classical'' propagator~(\ref{Path21}) employing
relation~(\ref{Path12}), where Green function~(\ref{Path20}) for free
motion is used. As a result, we arrive at
\begin{eqnarray}\label{Path22}
\Pi_{\rm f}\left(X,\mu,\nu,X',\mu',\nu',t\right)&=&\frac {1}{4\,\pi^2}\int
k^2\,\frac {1}{2\,\pi t}\delta\left(y-z-k\nu'\right)\nonumber\\
&&\otimes\exp\left\{
\frac{i}{2t}\left(a+\frac {k\nu}{2}-y\right)
-\frac{i}{2t}\left(a-\frac {k\nu}{2}-z\right)\right.\nonumber\\
&&\qquad \left. +ik\left(X'-X+\mu a-\mu'\,\frac {yz}{2}\right)\right\}
\,dk\,dy\,dz\,da\,.
\end{eqnarray}
Integration in Eq.~(\ref{Path22}) over variable $a$ gives a delta-function,
namely,
$$\int da\Longrightarrow \delta \left(\frac {k\left(\nu+\mu t\right)
+z-y}{t}\right).$$
After introducing new variables
$$y+z=s\,,\qquad y-z=m$$
and integrating over variable $s,$ one obtains another delta-function
$$\int ds\Longrightarrow \delta \left(\frac {k\left(\nu+\mu' t\right)
+z-y}{t}\right).$$
After integrating first over variable $m$ and then over variable $k,$
one arrives at the expression
\begin{equation}\label{Path23}
\Pi_{\rm f}\left(X,\mu,\nu,X',\mu',\nu',t\right)=|t|\,
\delta\left(X-X'\right)\,\delta\left(\nu -\nu'+\mu t\right)
\,\delta\left(\nu-\nu'+\mu't\right).
\end{equation}
In view of the relationships for delta-functions,
\begin{eqnarray*}
\delta\left(x+y\right)\,\delta\left(x+z\right)&=&
\delta\left(x+y\right)\,\delta\left(y-z\right),\nonumber\\
\delta \left(\frac {x}{t}\right)&=&|t|\,\delta\left(x\right),\nonumber
\end{eqnarray*}
one obtains the result~(\ref{Path21}).

Another example is the harmonic oscillator, for which Green function 
has the form $\left(m=\omega=\hbar=1\right)$
\begin{equation}\label{Path24}
G_{\rm os}\left(x,y,t\right)=\frac {1}{\sqrt{2\,\pi i\sin t}}\,
\exp\left[
\frac {i}{2}\,\mbox{cot}\,t\left(x^2+y^2\right)
-\frac {ixy}{\sin t}\right].
\end{equation}
By using formula~(\ref{Path12}), we obtain the ``classical'' propagator
for the harmonic oscillator,
\begin{eqnarray}\label{Path25}
\Pi_{\rm os}\left(X,\mu,\nu,X',\mu',\nu',t\right)&
=&\frac {1}{4\,\pi^2}\int \frac{k^2}{2\,\pi \,|\sin t|}
\,\delta\left(y-z-k\nu'\right)\nonumber\\
&&\otimes\exp\left\{
\frac {i}{2}\,\mbox{cot}\,t\left[\left(a+
\frac {k\nu}{2}\right)^2+y^2\right]
-\frac {iy}{\sin t}\left(a+
\frac {k\nu}{2}\right)\right.\nonumber\\
&&\qquad \left.-\frac {i}{2}\,\mbox{cot}\,t\left[\left(a-
\frac {k\nu}{2}\right)^2+z^2\right]\right.
-\frac {iz}{\sin t}\left(a-
\frac {k\nu}{2}\right)\nonumber\\
&&\qquad \left.-ik\left(X-X'\right)
-i\mu' k\,\frac {y+z}{2}+ik\mu a\right\}\,dk\,dy\,dz\,da\,.
\end{eqnarray}
Integration in Eq.~(\ref{Path25}) over variable $a$ gives a delta-function,
namely,
$$\int da\Longrightarrow \delta \left(k\left(\nu 
\,\mbox{cot}\,t
+\mu \right)-\frac{y-z}{\sin t}\right).$$
Integration in Eq.~(\ref{Path25}) over variable $s=y+z$ results in
another delta-function
$$\int ds\Longrightarrow \delta \left(\left(y-z\right)\,\mbox{cot}\,t
-k\mu'-\frac{k\nu}{\sin t}\right).$$
Integration over variable $m=y-z$ gives the term
\begin{equation}\label{Path26}
\int dm\Longrightarrow \frac {k^2}{|\sin t|}
\delta\left(k\left(\nu 
\,\mbox{cot}\,t+\mu\right)-\frac {k\nu'}{\sin t}\right)\,
\delta\left(k\nu' 
\,\mbox{cot}\,t-k\mu'-\frac {k\nu}{\sin t}\right).
\end{equation}
Employing properties of delta-function, after the last integration over
variable $k$, one obtains
\begin{eqnarray}\label{Path27}
\Pi_{\rm os}\left(X,\mu,\nu,X',\mu',\nu',t\right)&
=&|\sin t|\,
\delta\left(X-X'\right)
\,\delta \left(\nu \cos t+\mu \sin t-\nu'\right)
\,\delta \left(\nu' \cos t-\mu' \sin t-\nu\right)\nonumber\\
&=&\delta\left(X-X'\right)
\,\delta \left(\nu \cos t+\mu \sin t-\nu'\right)
\,\delta \left(\mu \cos t-\nu \sin t-\mu'\right).
\end{eqnarray}
Expression~(\ref{Path27}) is the ``classical propagator'' for the harmonic
oscillator; it was derived using the different technique 
in~\cite{ManciniFP,OlgaVolJRLR97,RosaPhR98}.

\section{Quasiclassical Approximation for the ``Classical'' Propagator}

\noindent

 Formula~(\ref{Path13}) gives exact expression for the ``classical''
propagator in terms of the quantum Green function
$G\left(x_2,\,x_1,\,t\right).$
On the other hand,
there exists the quasiclassical van-Fleck formula for Green function of
the Schr\"odinger equation
\begin{equation}\label{Path30}
G^{(q)}\left(x_2,x_1,t\right)
=\frac {C(t)}{\sqrt{|{\cal D}_x|}}\,
e^{iS\left(x_2,\,x_1,\,t\right)},
\end{equation}
where 
$${\cal D}_x=\left|\begin{array}{clcr}
\displaystyle{\frac{\partial^2S}{\partial x_1^2}}&
\displaystyle{\frac{\partial^2S}{\partial x_1\partial x_2}}\\[4mm]
\displaystyle{\frac{\partial^2S}{\partial x_2\partial x_1}}&
\displaystyle{\frac {\partial^2S}{\partial x_2^2}}\end{array}\right|.$$
In formula~(\ref{Path30}), 
$S\left(x_2,\,x_1,\,t\right)$ 
is the classical action,
satisfying the classical Hamilton equation
\begin{equation}\label{Path31}
\frac {\partial S}{\partial t}
+{\cal H}\left(p,x_2\right)
|_{p\rightarrow \partial S/\partial x_2}=0
\end{equation}
for the classical system with the Hamiltonian ${\cal H}\left(p,\,x\right)$,
and the function $C(t)$ is taken according to the Schr\"odinger equation
for Green function.
 
One can use the quasiclassical approximation~(\ref{Path30}) for Green function
in order to obtain the ``classical'' propagator in the quasiclassical
approximation. We arrive at 
\begin{eqnarray}\label{Path32}
\Pi^{(q)}\left(X,\mu,\nu,X',\mu',\nu',t\right)&=&\frac {1}{4\,\pi^2}
\int k^2\,\delta\left(y-z-k\nu'\right)\frac {|C(t)|^2}
{\sqrt{|{\cal D}_x{\cal D}_y|}}\nonumber\\
&&\otimes \exp\left\{iS\left(a+\frac {k\nu}{2},\,y,\,t\right)
-iS\left(a-\frac {k\nu}{2},\,z,\,t\right)\right.\nonumber\\
&&\qquad \left. +ik\left[\left(X'-X\right)+\mu a-\mu'\,\frac {y+z}{2}
\right]\right\}\,dk\,dy\,dz\,da\,,
\end{eqnarray}
where the notation
$$x_2=a+\frac {k\nu}{2}\,,\qquad x_1=y\,,\qquad
y_1=a-\frac {k\nu}{2}\,,\qquad y_2=z$$
is used.

Quasiclassical formula~(\ref{Path32}) is the result of calculations
of the ``classical'' propagator in terms of path integral~(\ref{Path19})
for the functional of the classical action of the form
\begin{equation}\label{Path33}
S\left[x(t)\right]=S\left[\underline {x}(t)\right]+
\left(S\left[x(t)\right]-
S\left[\underline {x}(t)\right]\right),
\end{equation}
where $\underline {x}(t)$ is the classical trajectory.

Then we used series expansion for the difference of the functional and 
its value at the classical trajectory. For linear systems with quadratic 
Hamiltonians, series~(\ref{Path33}) contains only two terms and the exact
result for the ``classical'' propagator coincides with the result of
the quasiclassical approximation.

\section{Optical Tomography in Bargmann Representation}

\noindent 

In previous sections, we used the symplectic tomography approach.
In the optical tomography approach~\cite{VogRis}, the Wigner function
is reconstructed, if one considers the marginal distribution
$w\left(X,\,\varphi,\,t\right)$
of the homodyne observable $X$,
the distribution being dependent on the rotation angle $\varphi.$
Till now, the evolution equation was obtained for the marginal distribution
$w\left(X,\,\mu,\,\nu,\,t\right)$
in the symplectic tomography approach.
Since
\begin{equation}\label{Path40}
w\left(X,\varphi,t\right)=w\left(X,\cos \varphi,\sin \varphi,t\right),
\end{equation}
the possibility to derive the evolution equation for the optical
marginal distribution arises. To do this, let us introduce complex
variables
\begin{equation}\label{Path41}
 z=\mu +i\nu\,;\qquad \bar{z}=\mu -i\nu\,,
\end{equation}
which are similar to the variables used in the Bargmann representation
of coherent states.

The inverse of Eq.~(\ref{Path41}) reads
\begin{equation}\label{Path42}
 \mu =\frac {z+\bar z}{2}\,;\qquad \nu=\frac {z-\bar z}{2}\,.
\end{equation}
The evolution equation for the marginal probability distribution
has the appearance~\cite{ManciniPL,ManciniFP}
\begin{equation}\label{Path43}
\dot  w-\mu\,\frac{\partial}{\partial \nu}\,w
-i\left[V\left(
-\frac{1}{\partial/\partial X}\,\frac{\partial}{\partial\mu}
-i\,\frac {\nu}{2}\,\frac {\partial}{\partial X}\right)
-V\left(-\frac{1}{\partial/\partial X}\,\frac {\partial}{\partial \mu}
+i\,\frac{\nu}{2}\,\frac{\partial}{\partial X}\right)\right]w=0\,.
\end{equation}
Using for variables the notation determined by~(\ref{Path41}), 
we arrive at
\begin{equation}\label{Path44}
\dot  w-i\,\frac{z+\bar z}{2}
\left(\frac{\partial}{\partial z}-
\frac{\partial}{\partial \bar z}\right)
w-i\left\{V\left[
-\frac{1}{\partial/\partial X}
\left(\frac{\partial}{\partial z}-
\frac{\partial}{\partial \bar z}\right)-\frac{1}{4}\,
\left(z-\bar z\right)\frac{\partial}{\partial X}\right]
-\mbox {c.c.}\right\}w=0\,.
\end{equation}
If one makes the substitution
$$
z\Longrightarrow e^{i\varphi},\,\qquad
\bar z \Longrightarrow e^{-i\varphi},
$$
$w\left(X,\,z,\,\bar z,\,t\right)$,
being the solution to Eq.~(\ref{Path44}),
is the marginal distribution of the optical tomography approach,
namely,
\begin{equation}\label{Path45}
w\left(X,\varphi,t\right)=w\left(X,e^{i\varphi},e^{-i\varphi},
t\right).
\end{equation}

Thus, if one obtains the solution to the evolution equation for the marginal
distribution in the Bargmann representation, the marginal distribution  
of the symplectic tomography approach can be derived by means of 
Eq.~(\ref{Path45}), as well. Analogous substitutions can be applied for the
``classical'' propagator and its path integral representation.

\section{Conclusion}

\noindent

 Formulas~(\ref{Path12}) and (\ref{Path13}), which give the expression 
for the ``classical'' propagator (determining the evolution of the quantum 
system in the probability representation of quantum mechanics) in terms of 
the quantum propagator (Green function of the Schr\"odinger evolution 
equation) are the main result of this study.

Another important result is given by formula~(\ref{Path19}) which desribes
the ``classical'' propagator in terms of path integral determined by the 
functional of the classical action. Formula~(\ref{Path13}) obtained in this
study is the inverse of the expression of the quantum propagator in terms 
of the ``classical'' propagator which was found in~\cite{OlgaVolJRLR97}
(see also~\cite{RosaPhR98}\,).

Thus, the invertable map of density matrix onto marginal probability
distribution, which determines completely the quantum state, is accomplished
by the invertable map of the quantum propagator onto the ``classical''
propagator, which describes completely the evolution of the quantum system.

The discussed properties of the ``classical'' propagator can be used for 
studying evolution of states in the framework of the approach considered
in~\cite{OLGA}.

\section*{Acknowledgments}

\noindent

 This study was partially supported by the Russian Foundation for Basic
Research.

\end{document}